\documentclass[11pt,fleqn]{article}
\usepackage{amssymb,latexsym,amsmath,amsfonts,graphicx}

    \advance\textheight by \topskip

    \numberwithin{equation}{section}

    \def\Re{{\rm Re \,}}
    \def\Im{{\rm Im \,}}
    \def\Ai{{\rm Ai \,}}

    \def\bigO{{\cal O}}
    
    \def\Res{{\rm Res}}

\newcommand{\e}{\epsilon}
\newcommand{\lb}{\lambda}

    \newtheorem{theorem}{Theorem}[section]

    \newtheorem{proposition}[theorem]{Proposition}
    
    \newtheorem{Definition}[theorem]{Definition}
    
    \newtheorem{Remark}[theorem]{Remark}
    \newenvironment{remark}{\begin{Remark}\rm}{\end{Remark}}
    \newtheorem{Example}[theorem]{Example}
    
    \newtheorem{Assumptions}[theorem]{Assumptions}

    \newenvironment{proof}%
    {\rm \trivlist \item[\hskip \labelsep{\bf Proof. }]}%
    {\hspace*{\fill}$\Box$\endtrivlist}
    {\rm \trivlist \item[\hskip \labelsep{\bf Proof}]}%
    {\hspace*{\fill}$\Box$\endtrivlist}

\begin{document}
\title{Painlev\'e II asymptotics near the leading edge of the oscillatory zone for the Korteweg-de Vries equation in the small dispersion limit}
\author{T. Claeys and T. Grava}

\maketitle

\begin{abstract}
In the small dispersion limit, solutions to the Korteweg-de Vries equation develop an interval of fast oscillations after a certain time. We obtain a universal asymptotic expansion for the Korteweg-de Vries solution near the leading edge of the oscillatory zone up to second order corrections. This expansion involves the Hastings-McLeod solution of the Painlev\'e II equation. We prove our results using the Riemann-Hilbert approach.
\end{abstract}

\section{Introduction and statement of result}
In recent works, critical behavior of solutions to Hamiltonian perturbations of hyperbolic and elliptic systems has been studied. In particular the universal nature of this critical behavior is remarkable. For hyperbolic one- and two-component systems as well as for elliptic two-component systems, it has been conjectured by Dubrovin et al.\ \cite{Dubrovin, Dubrovin1, DubrovinGravaKlein} that solutions can be described in a certain critical regime by special solutions to the Painlev\'e I equation and its hierarchy. For several hyperbolic systems, two other critical regimes have been observed. One of those regimes, which is related to the second Painlev\'e equation, is the subject of the present paper.

\medskip

\begin{figure}[!htb]
\centering
\includegraphics[width=5in]{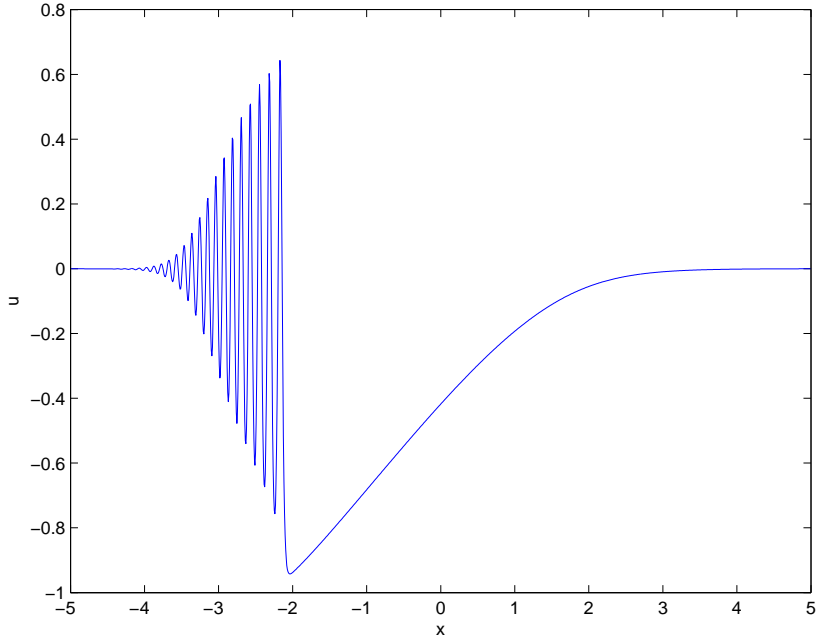}
\caption{Solution of the KdV equation with $\e=10^{-2}$, initial data $u_0(x)=-\mbox{sech}^2(x)$, and $t=0.4$. }
\label{fig1}
\end{figure}

As a prototype-example of a Hamiltonian perturbation of a hyperbolic one-com\-po\-nent PDE, we consider the small dispersion limit where $\e\to 0$ for the Korteweg-de Vries (KdV) equation
\begin{equation}
\label{KdV1}
u_t+6uu_x+\epsilon^2u_{xxx}=0,\quad x\in\mathbb{R},\;t\in\mathbb{R}^+, \quad\e>0.
\end{equation}
We are interested in the KdV solution $u(x,t,\e)$ at time $t>0$ when starting at $t=0$ with initial data $u(x,0,\e)=u_0(x)$ which we assume to be negative, smooth, sufficiently decaying at $\pm\infty$, and with a single negative hump.
For a much wider class of initial data than the one we consider, it is known that the KdV solution exists at all positive times $t$.
Up to a certain time the KdV solution $u(x,t,\e)$ can be approximated by the solution to the Cauchy problem of the dispersionless equation
\begin{equation}
\label{hopf}
u_t+6uu_x=0,  \qquad u(x,0)=u_0(x),
\end{equation}
which is the Hopf equation.
It is a classical fact that the solution to the Hopf equation reaches a point of gradient catastrophe at a finite time $t_c>0$, where the derivative of the solution blows up. After this time, the Hopf solution does not remain well-defined and can only be continued as a multi-valued function. The dispersive term $\e^2u_{xxx}$ in the KdV equation (\ref{KdV1}) regularizes the gradient catastrophe for any $\e>0$. As a drawback, the KdV solution develops oscillations with wavelength $\bigO(\e)$ for $t>t_c$, see Figure~\ref{fig1}.

 Those oscillations were already described in the classical works of Lax and Levermore \cite{LL} and investigated in more detail by Venakides \cite{V2}, Tian \cite{FRT}, and Deift, Venakides, and Zhou \cite{DVZ, DVZ2}. Under mild assumptions on the initial data, $u(x,t,\e)$ can be asymptotically described using Whitham equations \cite{W} and elliptic $\theta$-functions. Up to a fixed time after the time of gradient catastrophe, say for $t_c<t<T$, there is one oscillatory interval $[x^-(t), x^+(t)]$. For $x\in(x^-(t), x^+(t))$ (and $x$ fixed), we then have
\begin{multline}\label{elliptic}
u(x,t,\e)\approx \beta_1+\beta_2+\beta_3+2\alpha\\
+2\e^2\frac{\partial^2}{\partial x^2}\log\vartheta\left(\frac{\sqrt{\beta_1-\beta_3}}{2\e K(s)}[x-2t(\beta_1+\beta_2+\beta_3)-q];\tau\right).
\end{multline}
Here $\beta_1(x,t), \beta_2(x,t), \beta_3(x,t)$ evolve according to the Whitham equations, and
\begin{equation}
\alpha=-\beta_1+(\beta_1-\beta_3)\frac{E(s)}{K(s)}, \quad \tau=i\frac{K'(s)}{K(s)},\quad s^2=\frac{\beta_2-\beta_3}{\beta_1-\beta_3},
\end{equation}
where $K(s)$ and $E(s)$ are the complete elliptic integrals of the first and second kind, $K'(s)=K(\sqrt{1-s^2})$, and $\vartheta(z;\tau)$ is the Jacobi elliptic theta function given by the Fourier series
\[\vartheta(z;\tau)=\sum_{n\in\mathbb Z}e^{\pi i n^2\tau+2\pi i n z}.\]
For $t>t_c$ but $x$ bounded away from the oscillatory interval $[x^-(t), x^+(t)]$, the asymptotic behavior of the KdV solution is still determined by the Hopf equation.

\medskip

\begin{figure}[t]\label{Figure: cusp}
\begin{center}
\includegraphics[scale=0.35]{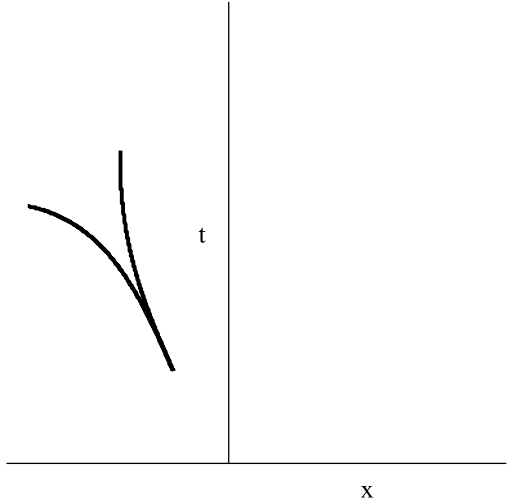}
\end{center}
\caption{Cusp-shaped oscillatory region in the $(x,t)$-plane for $t>t_c$.}
\end{figure}

In the $(x,t)$-plane up to a certain time $T>t_c$, we distinguish a 'Hopf region' and an oscillatory region which is cusp-shaped. The caustics separating the two regions consist of the leading edge, the trailing edge, and the point of gradient catastrophe itself, see Figure \ref{Figure: cusp}. Although we do not use this fact, we note that those borders are the curves on which the so-called Lax-Levermore maximizer \cite{LL, V2} is singular and the associated Riemann surface changes from genus $0$ to genus $1$.
  The transitional asymptotic description of the KdV solution $u(x,t,\e)$ for $(x,t)$ approaching the caustics, has escaped investigation for a long time.
Near the point and time of gradient catastrophe, it was proved recently that the KdV solution can be approximated using a higher order Painlev\'e I transcendent \cite{CG}. This supports the conjecture of Dubrovin \cite{Dubrovin}, who claimed that this Painlev\'e I-type behavior should be to a large extent universal, see also \cite{Dubrovin1}. The behavior near the trailing edge has not been clearly understood so far, but we intend to come back to this problem in future work. Near the leading edge, multiple scale analysis and numerical results by Grava and Klein \cite{GK2} indicate that the Hastings-McLeod solution to the Painlev\'e II equation takes over the role of the higher order Painlev\'e I transcendent. It is the aim of this paper to prove this rigorously for a large class of initial data. It is expected that the Hastings-McLeod solution to Painlev\'e II arises universally near the leading edge for a whole class of perturbations to hyperbolic equations, including (among others) the Camassa-Holm equation \cite{CH, GK1} and the de-focusing nonlinear Schr\"odinger equation.

\medskip

It is remarkable that the KdV solutions can be approximated by Painlev\'e equations in the critical regimes described above. Those Painlev\'e equations have appeared in many branches of pure and applied mathematics during the last decades, for an overview we refer to \cite{FIKN}.
An application of particular interest for us is the appearance of the first and second Painlev\'e equation in the theory of random matrices, see e.g.\ \cite{FIK, DK, BI, CK, CKV}.
In random matrix ensembles with certain unitary invariant probability measures, the eigenvalues accumulate on a finite union of intervals when the dimension of the matrices grows \cite{Deift, DKMVZ2, DKMVZ1}. Including a parametric dependence to the probability measure for the matrices, it is possible to observe transitional regimes where the number of intervals in the spectrum changes. This can happen in three different ways. The first one occurs when two intervals merge to a single interval, and this situation shows remarkable similarities with the leading edge for the KdV equation.
The second possibility, which should be related to the trailing edge for KdV, is that an interval in the spectrum shrinks and disappears afterwards.
Finally it could happen that the merging of two intervals takes place simultaneously with the disappearance of one of the intervals. This last phenomenon is related to the gradient catastrophe for KdV.

\medskip

The Painlev\'e II equation is the following second order ODE in the complex domain,
 \begin{equation}\label{PII}
 q''(s)=sq+2q^{3}(s).
\end{equation}
The special solution in which we are interested, is the Hastings-McLeod solution \cite{HastingsMcLeod}
which is uniquely determined by the boundary conditions
\begin{align}
&\label{HM1}q(s)=\sqrt{-s/2}(1+o(1)),&\mbox{ as $s\to -\infty$,}\\
&\label{HM2}q(s)=\mbox{Ai}(s)(1+o(1)), &\mbox{ as $s\to +\infty$,}
\end{align}
where $\mbox{Ai}(s)$ is the Airy function.
Although any Painlev\'e II solution has an infinite number of poles in the complex plane, it is known \cite{HastingsMcLeod} that the Hastings-McLeod solution $q(s)$ is smooth for all real values of $s$.

\subsection{Statement of result}

As already mentioned, we consider negative smooth initial data with only one local minimum and with sufficient decay at $\pm\infty$.
To be more precise, we impose the following conditions.
 \begin{Assumptions}\label{assumptions}
\begin{itemize}
\item[(a)] $u_0(x)$ is real analytic and has an analytic continuation to the complex plane in a domain of the form
\[
\mathcal{S}=\{z\in\mathbb{C}: |\Im z|<\tan\theta_0|\Re z|\}\cup\{z\in\mathbb C: |\Im z|<\sigma\},
\]
for some $0<\theta_0<\pi/2$ and $\sigma>0$,
\item[(b)] $u_0(x)$ decays as $|x|\rightarrow\infty$ in $\mathcal{S}$ such that
\begin{equation}
u_0(x)=\bigO\left(\frac{1}{|x|^{3+d}}\right), \;\;d>0, \quad x\in \mathcal S,
\end{equation}
\item[(c)] $u_0(x)<0$ and $u_0$ has a single local minimum at the point
$x=0$, with
\[u_0'(0)=0, \qquad u_0''(0)>0,\]
and $u_0$ is normalized such
that $u_0(0)=-1$,
\end{itemize}
\end{Assumptions}
Condition (b) is needed in order to apply the scattering transform \cite{BDT} for the KdV equation. This transformation associates a reflection coefficient to the initial data $u_0$. Condition (a) yields certain analyticity properties of the reflection coefficient. The presence of only one local minimum as imposed in condition (c) is convenient in order to have the simplest possible situation in the study of the semiclassical asymptotics for the reflection coefficient, while the position of the minimum and the minimal value of $u_0$ can be chosen to be $0$ and $-1$, respectively, without loss of generality.

\medskip

The gradient catastrophe for the Hopf equation takes place at time
\begin{equation}\label{tc}
t_c=\dfrac{1}{\max_{\xi\in\mathbb{R}}[-6u'_0(\xi)]}.
\end{equation}
Let us denote $f_L$ for the inverse function of the decreasing part of the initial data $u_0$ (the part to the left of $0$), so that
\begin{equation}
\label{tc2}
t_c=-\frac{1}{6}\max_{\xi\in(-1,0)}[f_L'(\xi)]=-\frac{1}{6}f_L'(u_c).
\end{equation}
The gradient catastrophe is generic if $f_L'''(u_c)\neq 0$, and we will assume that this is the case for our initial data $u_0$.
It is known \cite{FRT, GT} that there exists a time $T>t_c$ such that for $t_c<t<T$, the leading edge $x^-(t)$ is determined uniquely by the system of
equations
\begin{align}
&\label{leading}x^-(t)=6tu(t)+f_L(u(t)),\\
&\label{leading2}6t+\theta(v(t);u(t))=0,\\
&\label{leading3}\partial_{v}\theta(v(t);u(t))=0,
\end{align}
with $u(t)>v(t)$ and with
\begin{equation}
\label{theta}
\theta(v;u)=\dfrac{1}{2\sqrt{u-v}}\int_{v}^u\dfrac{f'_L(\xi)d\xi}{\sqrt{\xi-v}}.
\end{equation}
Furthermore $x^-(t)$, $u(t)$, and $v(t)$ are smooth functions of $t$.
Throughout the rest of the paper, whenever we refer to $u$, we mean by this the solution of the system (\ref{leading})-(\ref{leading3}) for a given time $t_c<t<T$, while we denote the solution of the KdV equation as $u(x,t,\e)$ and the solution of the Hopf equation as $u(x,t)$.
The equations  (\ref{leading})-(\ref{leading3}) correspond to the Whitham equations in the confluent case where
\[\beta_1(x,t)=\beta_2(x,t)=v(t), \qquad \beta_3(t)=u(t),\]
see also (\ref{elliptic}).
Note that the point of gradient catastrophe corresponds to $\beta_1(x,t)=\beta_2(x,t)=\beta_3(x,t)$ (and thus $u(t)=v(t)$) and the trailing edge to the case where $\beta_1(x,t)<\beta_2(x,t)=\beta_3(x,t)$.

\medskip

We will prove the following result.
\begin{theorem}\label{theorem: main}
Let $u_0(x)$ be initial data for the Cauchy problem of the KdV equation
satisfying the conditions given in Assumptions \ref{assumptions}, and furthermore such that the generic condition $f_L'''(u_c)\neq 0$ holds with $u_c$ given by (\ref{tc2}). Then there exists a time $T>t_c$ such that the following is true for any fixed $t$ such that $t_c<t<T$.
Take a double scaling limit where we let $\epsilon\to
0$ and at the same time we let $x\to x^-(t)$  in such a way that
\begin{equation}
\lim \dfrac{x- x^-(t)}{\e^{2/3}}=X \ \in\mathbb R.
\end{equation}
In this double scaling limit, the solution $u(x,t,\epsilon)$ of the KdV equation (\ref{KdV1}) with initial data $u_0$ has the following asymptotic expansion,
\begin{multline}
\label{expansionu}
u(x,t,\e) = u-\dfrac{4\e^{1/3}}{c^{1/3}}q\left[s(x,t,\e)\right]
\cos\left(\frac{\Theta(x,t)}{\epsilon}+\e^{1/3}\Theta_1(x,t,\e)\right)\\+\frac{x-x^-}{6t+f'_L(u)}
-\dfrac{4\e^{2/3}} {c^{2/3}(u-v)} q\left[s(x,t,\e)\right]^2
\sin^2\left(\frac{\Theta(x,t)}{\epsilon}\right)+O(\e).
\end{multline}
Here $x^-$ and $v<u$ (each of them depending on $t$) solve the system (\ref{leading}), and the phase $\Theta(x,t)$ is given by
\begin{equation}
\label{Theta}
\Theta(x,t)=2\sqrt{u-v}(x-x^-)+2\int_{v}^{u}(f_L'(\xi)+6t)\sqrt{\xi-v}d\xi.
\end{equation}
Furthermore
\begin{equation}\label{def c}
c=-\sqrt{u-v}\dfrac{\partial^2}{\partial v^2}\theta(v;u)>0,\qquad s(x,t,\e)=-\frac{x-x^-}{c^{1/3}\sqrt{u-v}\,\e^{2/3}},
\end{equation}
with $\theta$ defined by (\ref{theta}), and $q$ is the Hastings-McLeod solution to the Painlev\'e II equation.
The correction to the phase $\Theta_1(x,t,\e)$ takes the form
\begin{multline}
\label{Theta1}
\Theta_1(x,t,\e)= \frac{1}{c^{1/3}}\left[\left(\frac{q'}{q}+p\right)\frac{\partial^3_{v^3}\theta(v;u)}{6\partial^2_{v^2}\theta(v;u)}-\frac{5p+\frac{q'}{q}}{4(u-v)}\right.
\\ \left.+\dfrac{s(x,t,\e)^2}{4}\left(\frac{\partial^3_{v^3}\theta(v;u)}{3\partial^2_{v^2}\theta(v;u)}-\frac{3}{2(u-v)}
+\dfrac{2c\sqrt{u-v}}{6t+f'_L(u)}
\right)\right],
\end{multline} where we used the notations
\[q=q(s), \qquad q'=q'(s),\qquad p=p(s)=-q^4(s)-sq^2(s)+q'(s)^2,\] with $s=s(x,t,\e)$.
\end{theorem}
\begin{remark}
Note that the leading order term in the expansion (\ref{expansionu}) of $u(x,t,\e)$ is given by $u(t)$. The second term in (\ref{expansionu}) is of order $\e^{1/3}$, while the further terms are of order $\e^{2/3}$.
From the $\bigO(\e^{1/3})$-term, we observe that $u(x,t,\e)$ develops oscillations of wavelength $\bigO(\e)$ at the leading edge. Zooming in at the leading edge, the amplitude of the oscillations is of order $\e^{1/3}$ and proportional to the Hastings-McLeod solution $q$. If we formally take the limit $X\to -\infty$ (so that $x$ lies to the left of the leading edge), the terms with the oscillations disappear due to the exponential decay of $q$, see (\ref{HM2}). We are then left with only two terms in (\ref{expansionu}), which are the first two terms in the Taylor series of the Hopf solution $u(x,t)$ near $x^-$. This is not surprising since we know that $u(x,t,\e)$ is approximately equal to $u(x,t)$ outside the oscillatory interval.
\end{remark}
\begin{remark}
The expansion (\ref{expansionu}) is in agreement with the numerical results obtained in \cite{GK2}. There is also a remarkable similarity between (\ref{expansionu}) and an asymptotic expansion obtained in \cite{BI, CKV} for the recurrence coefficients of orthogonal polynomials with respect to critical exponential weights. Those polynomials describe eigenvalue statistics in unitary random matrix ensembles. This can be interpreted as one instance of the more general analogy between Hamiltonian perturbations of hyperbolic equations and random matrices, which we discussed before.
\end{remark}
\begin{remark}
The time $T$ in Theorem \ref{theorem: main} should be chosen in such a way that the system (\ref{leading})-(\ref{leading3}) is solvable for $t_c<t<T$, and furthermore in such a way that Proposition \ref{prop phi} below holds true for $t_c<t<T$. We do not go into further detail here, but we note for the interested reader that these conditions are equivalent to the regularity of the Lax-Levermore maximizer for $t_c<t<T$ and $x=x^-(t)$.
\end{remark}
\begin{remark}
It is unclear to what extent our result remains valid if the gradient catastrophe is not generic, i.e.\ $f_L'''(u_c)=0$. For special choices of non-generic initial data, it is likely that the KdV asymptotics are described by higher members of the Painlev\'e II hierarchy.
\end{remark}

The proof of our theorem relies on a Riemann-Hilbert (RH) problem which characterizes solutions to the KdV equation. This RH problem can be constructed using the direct scattering transform, while solving the RH problem corresponds to the inverse scattering transform.
First of all in Section \ref{section: inverse scattering}, we construct the RH problem and we collect some known asymptotic results about the reflection coefficient which is associated to the initial data $u_0$ via the direct scattering transform. Using those asymptotics, we can start performing the Deift/Zhou steepest descent method \cite{DZ1, DVZ, DVZ2} in order to find asymptotics for the relevant RH problem. This is the content of Section \ref{section: RH}, where we will apply a number of transformations to the RH problem in order to have suitable jump conditions. After a contour deformation, we will obtain a RH problem with jump matrices which decay uniformly to constant matrices when $\e\to 0$, except near the special points $u$ and $v$. We will then construct an outside parametrix away from $u$ and $v$, and two local parametrices, one near $u$ and one near $v$. The local parametrix near $u$ will be constructed using the Airy function, and the one near $v$ using special functions related to the Painlev\'e II equation.
After matching the three parametrices, we obtain uniform asymptotics for the solution to our RH problem in Section \ref{section: final}, and consequently also for the KdV solution $u(x,t,\e)$ when $x$ approaches the leading edge of the oscillatory zone at an appropriate rate.

\section{RH problem for KdV and asymptotics for the reflection coefficient}\label{section: inverse scattering}

In this section we state the RH problem for the KdV equation, and we give a number of asymptotic results concerning the reflection coefficient for the Schr\"odinger equation with potential $u_0$. The RH problem has been constructed in \cite{Shabat, BDT}, and will be the starting point of our asymptotic analysis later on. The asymptotic results for the reflection coefficient were obtained in \cite{Ramond} and are crucial in order to perform a rigorous Deift/Zhou steepest descent analysis of the RH problem. A somewhat more detailed construction of the RH problem and the reflection coefficient can also be found in \cite{CG}.

\subsection{Construction of the RH problem}
The KdV equation (\ref{KdV1}) can be written in the Lax form using the operators
\[
L=\epsilon^2\dfrac{d^2}{dx^2}+u(x,t,\e),\quad
A=4\epsilon^2\dfrac{d^3}{dx^3}+3\left(u(x,t,\e)\dfrac{d}{dx}+\dfrac{d}{dx}u(x,t,\e)\right).
\]
The KdV equation can then indeed be written as
\begin{equation}
\label{Lax} \dot{L}=[L,A],
\end{equation}
where $\dot{L}=\dfrac{dL}{dt}$ is the operator of multiplication
by $u_t(x,t,\e)$. A remarkable consequence of this Lax form is that the spectrum of $L$ does not depend on time if $u(x,t,\e)$ evolves according to the KdV equation \cite{GGKM}. For negative initial data $u_0$, it is known that
the Schr\"odinger operator $L$ with potential $u_0$ has no point spectrum \cite{Ramond}.

\medskip

  Another important consequence of the Lax form (\ref{Lax}) for the KdV equation is that the solution $u(x,t,\e)$ can (for any $x, t, \e$) be recovered from a RH boundary value problem, where one searches for a complex $2\times 2$  matrix-valued function $M$ which is analytic in $\mathbb C\setminus \mathbb R$ and for which the boundary values on $\mathbb R$ are connected in a prescribed way. Suppose that $M$ satisfies conditions of the following form.

\subsubsection*{RH problem for $M$}
\begin{itemize}
\item[(a)] $M(\lb;x,t,\e)$ is analytic for
$\lb\in\mathbb{C}\backslash \mathbb{R}$. \item[(b)] $M$ has continuous boundary values $M_+(\lb)$ and $M_-(\lb)$ when approaching $\lambda\in\mathbb R\setminus\{0\}$ from above and below, and
\begin{align*}&M_+(\lb)=M_-(\lb){\small \begin{pmatrix}1&r(\lb;\e)
e^{2i\alpha(\lambda;x,t)/\e}\\
-\bar r(\lb;\e)e^{-2i\alpha(\lb;x,t)/e}&1-|r(\lb;\e)|^2
\end{pmatrix}},&\mbox{ for $\lb<0$,}\\
&M_+(\lb)=M_-(\lb)\sigma_1,\quad
\sigma_1=\begin{pmatrix}0&1\\1&0\end{pmatrix},&\mbox{ for $\lb>0$},
\end{align*}
with $\alpha$ given by
\begin{equation}\label{def alpha1}\alpha(\lambda;x,t)=4t(-\lambda)^{3/2}+x(-\lambda)^{1/2},\end{equation} defined with the branches of $(-\lambda)^{\alpha}$ which are analytic in $\mathbb C\setminus [0,+\infty)$ and positive for $\lambda<0$.
Furthermore for fixed $x,t,\e$, we impose $M(\lb)$ to be bounded near $0$.
\item[(c)] $M(\lb;x,t,\e)=\left(I+\bigO(\lambda^{-1})\right)
\begin{pmatrix}1&1\\&\\i\sqrt{-\lb}&-i\sqrt{-\lb}\end{pmatrix}\;\;$
as $\lb\rightarrow \infty$.
\end{itemize}

For a suitable function $r$ appearing in the jump matrices, the above RH problem is (uniquely) solvable, and the solution to the KdV equation with initial data $u_0$ can be recovered from the RH
problem by the formula
\begin{equation}\label{uM}
u(x,t,\e)=-2i\e\partial_x M_{1, 11}(x,t,\e),
\end{equation}
where
$M_{11}(\lb;x,t,\e)=1+\dfrac{M_{1, 11}(x,t,\e)}{\sqrt{-\lb}}+\bigO(1/\lb)$
as $\lb\to\infty$.
This is indeed known \cite{BDT, DVZ} to be the case if $r(\lambda;\e)$ is the reflection coefficient from the left for the Schr\"odinger equation with potential $u_0$,
\[L_0f=\lambda f, \qquad L_0=\epsilon^2\dfrac{d^2}{dx^2}+u_0(x),\] see
\cite{Faddeev, BDT} for the definition and basic results about the reflection coefficient. The RH solution $M$ can be expressed explicitly in terms of fundamental solutions to the eigenvalue equation $Lf=\lambda f$. As $L$ has no point spectrum, those fundamental solutions are not $L^2$-functions.

\subsection{Asymptotics for the reflection coefficient}

The reflection coefficient $r(\lambda;\e)$ is independent of $x$ (and $t$, since it is the reflection coefficient at time $0$) but depends on the initial data $u_0$ via the direct scattering transform.
Solving the Riemann-Hilbert problem corresponds to the inverse scattering transform. In general, there is no simple expression for $r(\lambda;\e)$ in terms of $u_0$. What is known, is that $r(\lambda;\e)$ is analytic in a sector of the form $\pi-2\theta_0\leq \arg\lambda\leq\pi$ (for some $\theta_0>0$) if Assumptions \ref{assumptions} are satisfied \cite{FujiieRamond}.
Furthermore a number of asymptotic results for $r(\lambda;\e)$ are known as $\e\to 0$.
For $\lambda<-1$, it is known that the reflection coefficient decays exponentially fast as $\e\to 0$. On the interval $(-1,0)$ on the other hand, $r(\lambda;\e)$ oscillates rapidly for small $\e$.
To be more precise, the following holds as $\e\to 0$,
\begin{align}
&r(\lambda;\e)= ie^{-\frac{2i}{\e}\rho(\lambda)}(1+\bigO(\e)),&\mbox{ for $\lambda\in(-1,0)$,}\\
&r(\lambda;e)=\bigO(e^{-\frac{c}{\e}}),&\mbox{ for $\lambda\in(-\infty,-1)$,}
\end{align}
where
$\rho(\lambda)$ is given by
\begin{equation}\label{def rho}
\rho(\lb)=f_L(\lambda)\sqrt{-\lambda}+\int_{-\infty}^{f_L(\lambda)}[\sqrt{u_0(x)-\lambda}-\sqrt{-\lambda}]dx,
\end{equation}
and where $f_L$ denotes, as before, the inverse of the decreasing part of the initial data $u_0$.

\medskip

We need some more precise estimates for the reflection coefficient, in particular near $-1$, where the transition takes place from oscillations to exponential decay.
Because of the assumptions we imposed on the initial data $u_0$, we can write
\begin{equation}
f_L(\lambda)=-g_L(\lambda)\sqrt{\lambda+1},
\end{equation}
where $g_L$ is real analytic in a neighborhood of the interval $[-1,0)$ and positive at $-1$, and where $\sqrt{\lambda +1}$ is analytic in $\mathbb C\setminus (-\infty,-1]$ and positive for $\lambda>-1$. We then have
\begin{equation}\label{def rho2}
\rho(\lb)=-g_L(\lambda)\sqrt{-\lambda(\lambda+1)}+\int_{-\infty}^{-g_L(\lambda)\sqrt{\lambda+1}}[\sqrt{u_0(x)-\lambda}-\sqrt{-\lambda}]dx.
\end{equation}
Now it is easy to see that $\rho$ is real analytic on $(-1,0)$ and that it has branch points at $-1$ and $0$. Observe that $\rho$ extends analytically to the region \begin{equation}\label{def Omega}\Omega_+:=\{\lb\in\mathbb C:-1-2\delta<\Re\lb<-2\delta, 0<\Im\lambda<2\delta\}\end{equation} for sufficiently small $\delta$.
Using the analyticity of the reflection coefficient in $\Omega_+$, one can thus define an analytic function $\kappa(\lambda)$ as follows,
\begin{equation}\label{def kappa}
\kappa(\lambda;\e)=-ir(\lb;\e)e^{\frac{2i}{\e}\rho(\lambda)}, \qquad\mbox{ for $\lb\in\Omega_+$}.
\end{equation}
In \cite{Ramond}, Ramond proved that, for any sufficiently small $\delta>0$, we have the following as $\e\to 0$,
\begin{align}
&\label{kappa1}\kappa(\lb;\e)=1+\bigO(\e),&\mbox{ uniformly for $\lb\in\Omega_+$ and $\Re\lb>-1+2\delta$,}\\
&\label{ref2}r(\lb;\e)=\bigO(e^{-\frac{c}{\e}}), &\mbox{ for $\lb<-1-2\delta$, $c>0$.}
\end{align}
In addition
\begin{equation}
\label{tau}
1-|r(\lambda;\epsilon)|^2\sim e^{-\frac{2}{\epsilon}\tau(\lambda)}, \qquad\mbox{ for $-1+2\delta<\lambda<0$,}
\end{equation}
where
\begin{equation}\label{deftau}
\tau(\lambda)=\int_{f_L(\lambda)}^{f_R(\lambda)}\sqrt{\lambda - u_0(x)}dx,
\end{equation}
and $f_R$ is the inverse of the increasing part of the initial data $u_0$ (the part to the right of the origin).
Near $-1$, say for $\lambda\in\Omega_+$ and $\Re\lambda<-1+2\delta$ (again for $\delta$ sufficiently small), another result in \cite{Ramond} states that
\begin{equation}
\kappa(\lambda;\e)= \dfrac{i}
{N\left(-\dfrac{i\tau(\lambda)}{\pi\epsilon}\right)}(1+O(\e)),\qquad\mbox{ as $\e\to 0$,}
\end{equation}
with $N$ given by
\begin{equation}
\label{N}
N(z)=\dfrac{\sqrt{2\pi}}{\Gamma(1/2+z)}e^{z\log(z/\e)},
\end{equation}
where $\Gamma$ is the Euler $\Gamma$-function.
Since $\tau$ maps the region $\{\lambda\in\Omega_+, \Re\lambda<-1+2\delta\}$ into the upper half of the complex plane, $\frac{-i\tau(\lambda)}{\pi\e}$ does not meet any poles of $\Gamma(1/2+z)$, and it follows that $\kappa(\lb;\e)$ is bounded for $\lambda\in\Omega_+$, $\Re\lambda<-1+2\delta$. Moreover, if we delete a small semi-disk of radius $\delta/2$ centered at $-1$ from $\Omega_+$,
it follows from the relation \[N(z)=1+\bigO(z^{-1}),\qquad \mbox{ as $z\to\infty$, }|\arg z|<\pi,\] together with (\ref{kappa1}) that
\begin{equation}\label{kappa3}
\kappa(\lambda;\e)=1+\bigO(\e), \qquad \mbox{ uniformly for $\lambda\in\Omega_+$, $|\lb+1|>\frac{\delta}{2}$, as $\e\to 0$.}
\end{equation}

\section{Asymptotic analysis of the RH problem near the leading edge}\label{section: RH}

From now on, we fix $t_c<t<T$ for sufficiently small $T>t_c$ and, when convenient, we suppress $t$-dependence in our notations.
We assume also that $|x-x^-|<\delta_0$ for some sufficiently small $\delta_0>0$.
Following the Deift/Zhou steepest descent method, we will apply a number of transformations to the RH problem for $M$. Here we follow similar lines as in \cite{DVZ, DVZ2}, where a similar analysis was carried out away from the caustics in the $(x,t)$-plane, and as in \cite{CG}, where a Riemann-Hilbert analysis was used near the point of gradient catastrophe. The most crucial new feature is the construction of local parametrices near the points $u$ and $v$.

\subsection{Construction of the $G$-function}

Define $G$ as follows
\begin{equation}
G(\lambda;x)=\frac{(u - \lambda)^{1/2}}{\pi}
\int_{u}^0\frac{\rho(\eta)-\alpha(\eta;x)}{\sqrt{\eta-u}(\eta
-\lambda)}d\eta,\label{def G}
\end{equation}
where the choice of $u$ is unimportant for now, as long as $u$ is independent of $x$ and $-1<u=u(t)<0$. Later on, it will be important that we choose $u$ according to (\ref{leading})-(\ref{leading3}). In (\ref{def G}),
$\rho$ is as in (\ref{def rho2}), and $\alpha$ is given as before in (\ref{def alpha1}),
\begin{equation}\label{def alpha}\alpha(\lb;x)=4t(-\lb)^{3/2}+x(-\lb)^{1/2}.\end{equation} The square root in front of the integral in (\ref{def G}) is chosen to be analytic in $\mathbb C\setminus [u,+\infty)$ and positive for $\lambda<u$, so that
$G$ is
analytic in $\mathbb C\setminus [u,+\infty)$ as well. $G$ has the following asymptotic behavior,
\begin{equation}\label{g0}G(\lambda;x)=G_1(x)\frac{1}{(-\lambda)^{1/2}}+\bigO(\lambda^{-1}),\qquad \mbox{ as $\lambda\to\infty$},\end{equation}
where $G_1$ is given by
\begin{eqnarray}\label{g1}
{G}_1(x)=\frac{1}{\pi}
\int_{u}^0\dfrac{\rho(\eta)-\alpha(\eta;x)}{\sqrt{\eta-u}}d\eta.
\end{eqnarray}
Note that $\rho(\eta)$ is independent of $x$, so that
\begin{equation}
\partial_x G_1(x)=-\frac{1}{\pi}\int_{u}^0\sqrt{\frac{-\eta}{\eta-u}}d\eta=\dfrac{u}{2}.
\label{dg1}
\end{equation}
Using Cauchy's theorem and keeping track of the branch cuts for the square roots, it is easily verified that $G$ has the properties
\begin{align}
&\label{Gjump1}G_+(\lb)+G_-(\lb)-2\rho(\lb)+2\alpha(\lb)=0, &\mbox{ as $\lb\in (u,0)$},\\
&\label{Gjump2}G_+(\lb)+G_-(\lb)=0, &\mbox{ as $\lb>0$}.
\end{align}

\subsection{Transformation $M\mapsto T$}

We define
\begin{equation}\label{def T}
T(\lambda;x,\e)=M(\lambda;x,\e)e^{-\frac{i}{\epsilon}
G(\lambda;x)\sigma_3},
\end{equation}
with $\sigma_3=\begin{pmatrix}1&0\\0&-1\end{pmatrix}$.
$T$ solves a RH problem with modified jump matrices compared to the ones for $M$. Using (\ref{def T}) and (\ref{Gjump1})-(\ref{Gjump2}),
we have the following RH problem for $T$.

\subsubsection*{RH problem for $T$}
\begin{itemize}
\item[(a)] $T$ is analytic in $\mathbb C\setminus \mathbb R$,
\item[(b)] $T_+(\lambda)=T_-(\lambda)v_T(\lambda)$, \ \ \ as
$\lambda\in\mathbb R$, with
\begin{equation}\label{vT}
v_T(\lambda)=\begin{cases}
\begin{array}{ll}
\sigma_1, &\mbox{ as $\lambda>0$,}\\[0.8ex]
\begin{pmatrix}1&r(\lb)e^{\frac{2i}{\e}(
G(\lb)+\alpha(\lb))}\\
-\bar r(\lb)e^{-\frac{2i}{\e}(G(\lb)+\alpha(\lb))} &1-|r(\lb)|^2
\end{pmatrix},&\mbox{ as $\lambda\in (-\infty,u)$,}\\[3ex]
\begin{pmatrix}e^{-\frac{i}{\e}(G_+(\lb)-G_-(\lb))}
&r(\lb)e^{\frac{2i}{\e}\rho(\lb)}\\
-\bar{r}(\lb)e^{-\frac{2i}{\e}\rho(\lb)}&(1-|r(\lb)|^2)e^{\frac{i}{\e}(
G_+(\lb)-G_-(\lb))}
\end{pmatrix},&\mbox{ as $\lambda\in (u,0)$,}
\end{array}
\end{cases}
\end{equation}
\item[(c)]
$T(\lambda)\sim\begin{pmatrix}1&1\\
i\sqrt{-\lambda}&-i\sqrt{-\lambda}\end{pmatrix}$ as
$\lambda\to\infty$.
\end{itemize}
Using (\ref{uM}) and (\ref{g0})-(\ref{dg1}), we obtain
\begin{equation}\label{uS1}
u(x,t,\e)=u-2i\e \partial_{x}T_{1,11}(x,t,\e),
\end{equation}
where
\begin{equation}
T_{11}(\lb;x,t,\e)=1+\frac{T_{1,11}(x,t,\e)}{\sqrt{-\lb}}+\bigO(\lb^{-1}), \qquad \mbox{ as $\lb\to\infty$.}
\end{equation}

\medskip

Let us now define $\phi(\lambda;x)$ by
\[
\phi(\lambda;x)={G}(\lambda;x)-\rho(\lambda)+\alpha(\lambda;x).
\]
This function is real for $\lb\in (-1,u)$ and it has branch points at $-1$ and $u$. Furthermore $\phi$ extends analytically to the region $\Omega_+$ given by (\ref{def Omega}).
By (\ref{Gjump1}) we have
\begin{equation}2\phi_+(\lambda;x)=G_+(\lambda;x)-G_-(\lambda;x), \qquad\mbox{ for $\lambda\in (u,0)$}.\label{phipmG}\end{equation}
The definition of $\phi$ is helpful because it enables us to write the jump matrices for $T$ in a more convenient form. Let us write $\phi^*(\lb)=\overline{\phi(\overline{\lb})}$ and $\kappa^*(\lb)=\overline{\kappa(\overline{\lb})}$, with $\kappa$ given by (\ref{def kappa}).
Using (\ref{vT}) together with (\ref{phipmG}) and reality of $\rho$ on $(-1,0)$, we obtain
\begin{equation}\label{vT2}
v_T(\lambda)=
\begin{pmatrix}e^{-\frac{2i}{\e}\phi_+(\lb;x)}
&i\kappa(\lambda;\e)\\
i\kappa^*(\lambda;\e)&(1-|r(\lb)|^2)e^{\frac{2i}{\e}\phi_+(\lb;x)}
\end{pmatrix},\qquad \mbox{ as $\lambda\in (u,0)$.}
\end{equation}
Furthermore we also have
\begin{equation}\label{vT3}
v_T(\lambda)=
\begin{pmatrix}1&i\kappa(\lb;\e)e^{\frac{2i}{\e}\phi(\lb;x)}\\
i\kappa^*(\lb;\e)e^{-\frac{2i}{\e}\phi^*(\lb;x)} &1-|r(\lb)|^2
\end{pmatrix},\qquad\mbox{ as $\lambda\in [-1-\delta,u)$,}
\end{equation}
where for $\lambda\in[-1-\delta,-1]$, $\phi(\lb)$, $\phi^*(\lb)$, $\kappa(\lb)$, and $\kappa^*(\lb)$ should be interpreted as the boundary values where $\Im\lambda\searrow 0$ (for the functions without star) and $\Im\lambda\nearrow 0$ (for the functions with star).
The formula (\ref{vT3}) for $\lambda\in(-1,u)$ follows directly from (\ref{vT}) and the reality of $\phi$. In order to obtain this formula as well for $[-1-\delta,-1]$, note that the $12$-entry of $v_T$ (given by (\ref{vT})) is analytic in $\Omega_+$, while the $21$-entry can be extended to an analytic function in $\overline{\Omega_+}$. The same holds true for the right hand side of (\ref{vT3}), and by the identity theorem, we obtain (\ref{vT3}) for $\lambda\in [-1-\delta,-1]$.

\medskip

If condition (\ref{leading}) holds, one verifies that (see \cite[Lemma 3.2]{CG})
\begin{equation}
\label{phi}
\phi(\lambda;x)=\sqrt{u-\lambda}(x-x^-)+\int_{\lambda}^{u}(f_L'(\xi)+6t)\sqrt{\xi-\lambda}d\xi.
\end{equation}
In addition we have
\begin{equation}\label{phiprime}
\phi'(\lambda;x)=-\dfrac{1}{2\sqrt{u-\lambda}}(x-x^-)-\sqrt{u-\lambda}[6t+\theta(\lambda;u)],
\end{equation}
where $\theta(\lambda;u)$ has been defined in (\ref{theta}).
If we let $u=u(t)$ and $v=v(t)$ be defined by (\ref{leading2})-(\ref{leading3}), it follows that
\begin{align}
&\label{phi1}
\phi'(v;x^-)=0,\qquad \phi''(v;x^-)=0,\\ &\label{phi2}\phi'''(v;x^-)=-\sqrt{u-v}\theta''(v;u)=:c>0.
\end{align}

\begin{proposition}\label{prop phi}
There exists a time $T>t_c$ such that for $t_c<t<T$, we have
\begin{equation}
\label{phiinequalities}
\begin{array}{ll}
\phi'(\lb;x^-)>0,&\mbox{ as $\lb\in[-1,u)\setminus\{v\}$,}\\[1.5ex]
\Im\phi_+(\lb;x^-)<0,&\mbox{ as $\lb\in (u,0]$,}\\[1.5ex]
-\tau(\lambda)+i\phi_+(\lb;x^-)<0,&\mbox{ as $\lb\in (u,0]$.}
\end{array}
\end{equation}
\end{proposition}
\begin{proof}
For $t=t_c$, it follows from (\ref{tc2}) that the function $f_L(\lb)+6t_c$ has a double zero at $\lambda=u=u_c$ (see also \cite{CG}) and that it is strictly negative on $[-1,0]\setminus\{u_c\}$. It follows from (\ref{theta}) and $u(t_c)=v(t_c)=u_c$ that the same holds for the function $6t_c+\theta(\lambda;u(t_c))$. For $t$ slightly bigger than $t_c$, because of the smoothness of $u(t)$, $6t+\theta(\lambda;u(t))$ can thus have at most two zeroes on $[-1,0]$. By (\ref{phi1}), we know that both of those zeroes must coincide at the point $v$. Consequently we have that $6t+\theta(\lambda;u(t))<0$ for $\lb\in[-1,0)\setminus\{v\}$ for sufficiently small times after the time of gradient catastrophe. This implies the first inequality. We also have $\Im\phi'_+(\lambda;x^-)<0$ for $\lb>u$, and together with $\phi(u;x^-)=0$, this implies the second inequality.

\medskip

For the last inequality, if $\tau$ is defined as in (\ref{deftau}), it is straightforward to verify that
\[-\tau(\lambda)+i\phi_+(\lb;x^-)=
-4t(\lambda-u)^{\frac{3}{2}}
+\int_{-1}^{u}\sqrt{\lambda-\xi}f'_L(\xi)d\xi
-\int_{-1}^{\lambda}\sqrt{\lambda-\xi}f'_R(\xi)d\xi,
\]
where  $f_R$ is the inverse function of the increasing part of the initial data $u_0(x)$.
The inequality follows directly because each of the three terms on the right hand side are negative.
\end{proof}
\begin{remark}
The smallest time $T$ for which the above inequalities do not hold any longer, is the first time after the gradient catastrophe at which the Lax-Levermore maximizer becomes singular for $x=x^-(t)$. Our RH analysis is only valid at times for which (\ref{phiinequalities}) holds.
\end{remark}

\subsection{Opening of the lens $T\mapsto S$}

\begin{figure}[t]
\begin{center}
    \setlength{\unitlength}{1.2mm}
    \begin{picture}(137.5,26)(22,11.5)
        \put(112,25){\thicklines\circle*{.8}}
        \put(45,25){\thicklines\circle*{.8}}
        \put(47,26){$-1-\delta$}
        \put(112,26){$0$}
        \put(90,25){\thicklines\circle*{.8}} \put(74,26){$v$}\put(90,26){$u$}
        \put(102,25){\thicklines\vector(1,0){.0001}}
        \put(90,25){\line(1,0){45}}
        \put(124,25){\thicklines\vector(1,0){.0001}}
        \put(22,25){\line(1,0){23}}
        \put(35,25){\thicklines\vector(1,0){.0001}}
        \qbezier(45,25)(60,45)(75,25) \put(61,35){\thicklines\vector(1,0){.0001}}
        \qbezier(45,25)(60,5)(75,25) \put(61,15){\thicklines\vector(1,0){.0001}}
\qbezier(75,25)(82.5,35)(90,25) \put(83,30){\thicklines\vector(1,0){.0001}}
        \qbezier(75,25)(82.5,15)(90,25) \put(83,20){\thicklines\vector(1,0){.0001}}
\end{picture}
\caption{The jump contour $\Sigma_S$ after opening of the lens}
\end{center}
\label{figure: lens}
\end{figure}
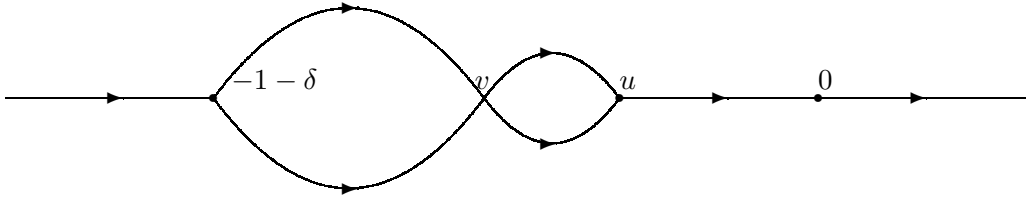

Note that, by (\ref{vT2}), we have the following factorization of the jump matrix for $T$,
\begin{equation}
v_T(\lb;x,\e)=\begin{pmatrix}1&0\\
i\kappa^*(\lb;\e)e^{-\frac{2i}{\e}
\phi_+^*(\lb;x)}&1
\end{pmatrix}\begin{pmatrix}1&i\kappa(\lb;\e)e^{\frac{2i}{\e}\phi_+(\lb;x)}\\
0&1
\end{pmatrix},\qquad\mbox{ for $\lambda\in (-1-\delta,u)$}.
\end{equation}
Because of this factorization, we can deform the jump contour to a lens-shaped contour as shown in Figure \ref{figure: lens}. Note that we start opening lenses at a point $-1-\delta$, and not at $-1$ itself. This is a modification compared to the analysis in \cite{DVZ, DVZ2} and is necessary to have jump matrices for $S$ which are uniformly close to constant matrices. We will determine the precise shape of the lenses later on.

\medskip

As $\phi_+$ can be extended analytically to $\Omega_+$ and $\phi_+^*$ to $\overline{\Omega_+}$, we can define $S$ as follows,
\begin{equation*}
S(\lambda)=\begin{cases}\begin{array}{ll}
T(\lambda)\begin{pmatrix}1&-i\kappa(\lb;\e)e^{\frac{2i}{\e}\phi(\lb;x)}\\
0&1
\end{pmatrix},&\mbox{ inside the lenses in the upper half plane,}\\[3ex]
T(\lambda)\begin{pmatrix}1&0\\
i\kappa^*(\lb;\e)e^{-\frac{2i}{\e}\phi^*(\lb;x)}&1
\end{pmatrix},&\mbox{ inside the lenses in the lower half plane},\\[3ex]
 T(\lambda), &\mbox{ elsewhere}.
\end{array}
\end{cases}
\end{equation*}

Now the RH problem for $S$ takes the following form.
\subsubsection*{RH problem for $S$}
\begin{itemize}
\item[(a)] $S$ is analytic in $\mathbb C\setminus \Sigma_S$,
\item[(b)] $S_+(\lambda)=S_-(\lambda)v_S$ for $\lambda\in\Sigma_S$,
with
\begin{equation}\label{vS}
v_S(\lambda)=\begin{cases}
\begin{array}{lr}
\begin{pmatrix}1& i\kappa(\lb)e^{\frac{2i}{\e}\phi(\lb)}\\
0&1
\end{pmatrix},&\mbox{ on $\Sigma_1$},\\[3ex]
 \begin{pmatrix}1&0\\
i\kappa^*(\lb;\e)e^{-\frac{2i}{\e}\phi^*(\lb;x)}&1
\end{pmatrix},&\mbox{ on $\Sigma_2$,}\\[3ex]
\begin{pmatrix}e^{-\frac{2i}{\e}\phi_+(\lb;x)}
&i\kappa(\lb;\e)\\
i\bar\kappa(\lb;\e)&(1-|r(\lb;\e)|^2)e^{\frac{2i}{\e}\phi_+(\lb;x)}
\end{pmatrix},&\mbox{ as $\lambda\in (u,0)$,}\\[3ex]
v_T(\lambda),&\mbox{\hspace{-3cm} as
$\lambda\in(-\infty,-1-\delta)\cup(0,+\infty)$.}
\end{array}
\end{cases}
\end{equation}
\item[(c)] $S(\lambda)\sim \begin{pmatrix}1&1\\
i\sqrt{-\lambda}&-i\sqrt{-\lambda}\end{pmatrix}$ as
$\lambda\to\infty$.
\end{itemize}

Since $T(\lb)=S(\lb)$ for large $\lb$, we have, using (\ref{uS1}),
\begin{equation}\label{uS}
u(x,t,\e)=u-2i\e \partial_{x}S_{1,11}(x,t,\e),
\end{equation}
where
\begin{equation}
S_{11}(\lb;x,t,\e)=1+\frac{S_{1,11}(x,t,\e)}{\sqrt{-\lb}}+\bigO(\lb^{-1}), \qquad \mbox{ as $\lb\to\infty$.}
\end{equation}
In the following proposition, we show that the jump matrices for $S$ tend uniformly to constant matrices as $\e\to 0$, except near the special points $u$ and $v$.
\begin{proposition}\label{prop jumps}
For any fixed neighborhoods $\mathcal U$ of $u$ and $\mathcal V$ of $v$, there exists $\delta_0>0$ such that for $|x-x^-|<\delta_0$, the jump matrices $v_S(\lambda)$ decay uniformly to the matrix $v^{(\infty)}(\lambda;x,\e)$ on $\Sigma_S\setminus (\mathcal U\cup \mathcal V\cup\{0\})$ if $\e\to 0$, with
\begin{equation}\label{vinfty}
v^{(\infty)}(\lambda)=\begin{cases}
\sigma_1,&\mbox{ on $(0, +\infty)$,}\\
i\sigma_1,&\mbox{ on $(u,0)$,}\\
I, &\mbox{elsewhere}.\end{cases}
\end{equation}
\end{proposition}
\begin{proof}
For $\lambda>0$, our claim is trivial. For $\lambda\leq -1-\delta$, we have that $v_S=v_T$ and using (\ref{vT}), exponential decay to the identity matrix simply follows from (\ref{ref2}) together with reality of $G$ and $\alpha$. \medskip
For $\lambda\in(u,0)\setminus \mathcal U$, if $x=x^-$, we have uniform exponential decay of the $11$-entry of $v_S$ (see (\ref{vS})) using the second inequality in Proposition \ref{prop phi} and of the $22$-entry using the third inequality in Proposition \ref{prop phi} together with (\ref{tau}). Furthermore if $x$ is close to $x^-$, the first term in (\ref{phi}) can only spoil the inequality in $\mathcal U$ and $\mathcal V$. The off-diagonal entries tend uniformly fast to $i$ using (\ref{kappa1}).

\medskip

Using the first equality in Proposition \ref{prop phi}, it follows from the Cauchy-Riemann conditions that, again for $x$ close to $x^-$, $\Im\phi(\lambda;x^-)>0$ on the upper parts of the lens if $\Re\lambda>-1+\delta$, and $\Im\phi^*(\lb;x^-)<0$ on the lower parts of the lens if $\Re\lambda>-1+\delta$, at least this is true if we choose the lenses sufficiently close to the real line. Again the inequalities remain valid for $x$ close to $x^-$ except in $\mathcal U$ and $\mathcal V$.

\medskip

Near $-1$, we have by (\ref{phi}) that
\[\Im\phi(\lb;x)=(x-x^-)\Im\sqrt{u-\lambda}+\Im\int_{\lb}^{-1}(f_L'(\xi)+6t)\sqrt{\xi-\lambda}d\xi.\] It is then straightforward to check that (for $x$ sufficiently close to $x^-$) $\Im\phi(\lambda;x)<0$ on the upper parts of the lens near $-1$, and $\Im\phi^*(\lb;x)>0$ on the lower parts of the lens near $-1$, including the point $-1-\delta$ itself. Together with (\ref{kappa3}), this implies convergence of $v_S$ on the lenses near $-1$ as well.
\end{proof}

\subsection{Outside parametrix}

Ignoring neighborhoods $\mathcal U$ and $\mathcal V$ of $u$ and $v$, and ignoring uniformly small jumps, we are left with the following RH problem.

\subsubsection*{RH problem for $P^{(\infty)}$}
\begin{itemize}
\item[(a)] $P^{(\infty)}:\mathbb C\setminus [u, +\infty) \to
\mathbb C^{2\times 2}$ is analytic, \item[(b)] $P^{(\infty)}$
satisfies the following jump conditions on $(u, +\infty)$,
\begin{align}
&P_+^{(\infty)}=P_-^{(\infty)}\sigma_1, &\mbox{ on $(0, +\infty)$},\\
&P_+^{(\infty)}= iP_-^{(\infty)}\sigma_1, &\mbox{ on $(u,0)$},\label{RHP Pinfty b}
\end{align}
\item[(c)]$P^{(\infty)}$ has the following behavior as $\lambda\to\infty$,
\begin{equation}P^{(\infty)}(\lambda)\sim
\begin{pmatrix}1&1\\ i(-\lambda)^{1/2} & -i(-\lambda)^{1/2}\end{pmatrix}
. \label{RHP Pinfty c}\end{equation}
\end{itemize}

This RH problem is solved by
\begin{equation}
\label{def Pinfty}
P^{(\infty)}(\lambda)=(-\lambda)^{1/4}(u-\lambda)^{-\sigma_3
/4}N, \qquad N=\begin{pmatrix}1&1\\ i& -i
\end{pmatrix}
\end{equation}
and we have the following behavior at infinity,
\begin{equation}\label{RHP Pinfty c2}P^{(\infty)}(\lambda)=\left(I+\frac{u}{4\lambda}\sigma_3+\bigO(\lambda^{-2})\right)
\begin{pmatrix}1&1\\ i(-\lambda)^{1/2} & -i(-\lambda)^{1/2}\end{pmatrix},
\quad\mbox{as $\lb\to\infty$.}\end{equation}
The outside parametrix will determine the leading order asymptotic behavior of $u(x,t,\e)$ in the double scaling limit. However, in order to do a rigorous asymptotic analysis and to obtain correction terms in the asymptotic expansion for $u(x,t,\e)$ (which is our main goal), we need to construct local parametrices in neighborhoods of the special points $u$ and $v$.

\subsection{Local parametrix near $u$ using the Airy function}
We will construct a parametrix $P_u$, in a disk $\mathcal U$ surrounding $u$, which has approximately (as $\e\to 0$) the same jumps as $S$ and which matches with the outside parametrix $P^{(\infty)}$ in the double scaling limit where $\e\to 0$ and simultaneously $x\to x^-$ in such a way that $x-x^-=\bigO(\e^{2/3})$. To be precise, the aim of this section is to construct a parametrix $P_u=P_u(\lambda;x,\e)$ satisfying the following conditions.
\subsubsection*{RH problem for $P_u$}
\begin{itemize}
\item[(a)] $P_u$ is analytic in $\overline{\mathcal U}\setminus\Sigma_S$.
\item[(b)] $P_u$ satisfies the jump conditions
\begin{align}
&P_{u,+}(\lb)=P_{u,-}(\lb)\begin{pmatrix}1&ie^{\frac{2i}{\e}\phi(\lb;x)}\\0&1\end{pmatrix},&\mbox{ as $\lb\in\Sigma_1\cap \mathcal U$},\\
&P_{u,+}(\lb)=P_{u,-}(\lb)\begin{pmatrix}1&0\\ie^{\frac{-2i}{\e}\phi^*(\lb;x)}&1\end{pmatrix}, &\mbox{ as $\lb\in\Sigma_2\cap \mathcal U$},\\
&P_{u,+}(\lb)=P_{u,-}(\lb)\begin{pmatrix}e^{\frac{-2i}{\e}\phi_+(\lb;x)}&i\\i&0\end{pmatrix}, &\mbox{ as $\lb>u$.}
\end{align}
\item[(c)] As $\e\to 0$ and $x\to x^-$ in such a way that $x-x^-=\bigO(\e^{2/3})$, $P_u$ matches with $P^{(\infty)}$ in the following way, \begin{equation}P_u(\lb)=P^{(\infty)}(\lb)\left(I+o(1)\right), \qquad\mbox{ for $\lb\in\partial \mathcal U$.}\end{equation}
\end{itemize}

\subsubsection{Model Airy RH problem}
We will use a model RH problem using the Airy function which is not precisely the same but very similar to the one used in \cite{Deift} and in many other works.
It involves functions $y_0, y_1, y_2$ solving the Airy differential equation and given by
\[
    y_j=y_j(\zeta)=e^{\frac{2\pi i j}{3}}\Ai(e^{\frac{2\pi ij}{3}}\zeta),\qquad j=0,1,2,
\]
 where $\Ai$ is the Airy function.
Let
\begin{align}
& A_1(\zeta)=-i\sqrt{2\pi}e^{-\frac{\pi i}{4}\sigma_3}
        \begin{pmatrix}
            -y_2 & -y_0\\
            -y_2' & -y_0'
        \end{pmatrix}e^{\frac{\pi i}{4}\sigma_3},\\[3ex]
& A_2(\zeta)=-i\sqrt{2\pi}e^{-\frac{\pi i}{4}\sigma_3}
        \begin{pmatrix}
            -y_2 & y_1\\
            -y_2' & y_1'
        \end{pmatrix}e^{\frac{\pi i}{4}\sigma_3},
        \\[3ex]
& A_3(\zeta)=-i\sqrt{2\pi}e^{-\frac{\pi i}{4}\sigma_3}
        \begin{pmatrix}
            y_0 & y_1\\
            y_0' & y_1'
        \end{pmatrix}e^{\frac{\pi i}{4}\sigma_3}.
\end{align}
Each of the functions $A_1$, $A_2$, and $A_3$ are analytic in the entire complex plane.
Using the well-known identity
$y_0+y_1+y_2=0$, it is easy to verify that
\begin{align}
            \label{RHP A: b1}
            & A_1(\zeta)=A_2(\zeta)
                \begin{pmatrix}
                    1 & i \\
                    0 & 1
                \end{pmatrix},
            \\[1ex]
            & A_2(\zeta)=A_3(\zeta)
                \begin{pmatrix}
                    1 & 0 \\
                    i & 1
                \end{pmatrix},
            \\[1ex]
            \label{RHP A: b3}
            & A_1(\zeta)=A_3(\zeta)
                \begin{pmatrix}
                    1 & i \\
                    i & 0
                \end{pmatrix}.
        \end{align}
Furthermore the asymptotic expansion at infinity for the Airy function shows that
\begin{align}\label{RHP:A-c}
            A_j(\zeta) &= e^{-\frac{\pi i}{4}\sigma_3}(-\zeta)^{-\frac{\sigma_3}{4}}N
                \left[I+\bigO\left(\zeta^{-3/2}\right)\right]
                e^{\frac{-2i}{3}(-\zeta)^{3/2}\sigma_3},
        \end{align}
        uniformly as $\zeta\to\infty$ in the sector $S_j$, $j=1, 2, 3,$ with
\begin{equation}
S_j=\{\zeta\in\mathbb C: \frac{j-1}{2}\pi<\arg\zeta<\frac{j+1}{2}\pi\}, \qquad j=1, 2, 3.
\end{equation}
The fractional powers in (\ref{RHP:A-c}) are principal branches, so $(-\zeta)^{\frac{k}{2}}$ is analytic for $0\leq\arg\zeta<2\pi$ and positive for $\arg\zeta=\pi$, and $N$ is given by (\ref{def Pinfty}).

\subsubsection{Construction of the parametrix}
We will now construct $P_u$ explicitly in terms of $A$.
Define, in the neighborhood $\mathcal U$ of $u$,  $\zeta(\lambda)$ such that
\begin{equation}\label{def hu}
\frac{-2}{3}\left(-\zeta(\lambda)\right)^{3/2}=\phi(\lambda;x^-).
\end{equation}
It then follows from (\ref{phi}) that $\zeta$ is a conformal mapping in a sufficiently small neighborhood $\mathcal U$ of $u$, with
\begin{equation}
\label{zu}
\zeta'(u)=\left(-f'_L(u)-6t\right)^{2/3}>0.
\end{equation}
Next define a function $s(\lambda;x)$ by
\begin{equation}\label{def su}
s(\lambda;x)(-\zeta(\lambda))^{1/2}=\phi(\lambda;x)-\phi(\lambda;x^-)=(x-x^-)\sqrt{u-\lambda},
\end{equation}
so that $s$ is analytic in $\overline{\mathcal U}$ with
\begin{equation}
\label{su}
s(u;x)=\left(-f'_L(u)-6t\right)^{-1/3}(x-x^-).
\end{equation}
Now we can define the parametrix $P$ as follows,
\begin{equation}\label{def Pu}
P_u(\lambda;x,\e)=E(\lambda;\e)A_j(\e^{-2/3}\zeta(\lambda)+\e^{-2/3}s(\lambda;x))e^{-\frac{i}{\e}\phi(\lb;x)\sigma_3},
\end{equation}
where $j=1$ in the region for $\lambda$ outside the lens in the upper half plane, $j=2$ inside the lens-shaped region, and $j=3$ outside the lens in the lower half plane.
The pre-factor $E$ given by
\begin{equation}
E(\lambda;\e)=P^{(\infty)}(\lambda)N^{-1}(-\e^{-2/3}\zeta(\lambda))^{\frac{1}{4}\sigma_3}e^{\frac{\pi i}{4}\sigma_3}.
\end{equation}
It is easily checked using (\ref{def Pinfty}) that $E$ is analytic in $\overline{\mathcal U}$.

\medskip

Using the jump conditions (\ref{RHP A: b1})-(\ref{RHP A: b3}) for $A$, it follows from (\ref{def Pu}) that $P_u$ indeed satisfies the required jump conditions. More subtle is the matching of $P_u$ with $P^{(\infty)}$. Let $\e\to 0$ and at the same time $x\to x^-$ in such a way that $x-x^-=\bigO(\e^{-2/3})$. Then also $s(\lambda;x)=\bigO(\e^{2/3})$, and we have by (\ref{def hu}) and (\ref{def su}),
\[-\frac{2i}{3}\left(-\e^{-2/3}\zeta(\lambda)-\e^{-2/3}s(\lambda)\right)^{3/2}=\frac{i}{\e}\phi(\lambda;x)-\frac{i}{4\e}\frac{s(\lambda;x)^2}{(-\zeta(\lambda))^{1/2}}
+\bigO(\e),
\qquad\mbox{ for $\lambda\in\partial \mathcal U.$}\]
Inserting the asymptotics (\ref{RHP:A-c}) for $A$ into (\ref{def Pu}) now yields the following matching of $P_u$ with $P^{(\infty)}$,
\begin{equation}\label{matching P u}
P_u(\lambda)=P^{(\infty)}(\lambda)\left[I-\frac{is(\lambda;x)^2}{4\e (-\zeta(\lambda))^{1/2}}\sigma_3-\frac{s(\lambda;x)}{4\zeta}\sigma_1+\bigO(\e)\right], \qquad\mbox{ for $\lambda\in\partial \mathcal U$,}
\end{equation}
in the double scaling limit.
This matching will contribute to the asymptotics for the KdV solution $u(x,t,\e)$ later on.

\subsection{Local parametrix near $v$ using $\Psi$-functions for Painlev\'e II}

We will now construct the local parametrix in a sufficiently small disk $\mathcal V$ surrounding $v$. This construction shows similarities with the construction of the local parametrices in \cite{CK, CKV}.
The parametrix $P_v=P_v(\lb;x,\e)$ should satisfy the following conditions.
\subsubsection*{RH problem for $P_v$}
\begin{itemize}
\item[(a)] $P_v$ is analytic in $\overline{\mathcal V}\setminus\Sigma_S$.
\item[(b)] $P_v$ satisfies the jump conditions
\begin{align}
\label{jump P1}&P_{v,+}(\lb)=P_{v,-}(\lb)\begin{pmatrix}1&ie^{\frac{2i}{\e}\phi(\lb;x)}\\0&1\end{pmatrix},&\mbox{as $\lb\in\Sigma_1\cap \mathcal V$},\\
\label{jump P3}&P_{v,+}(\lb)=P_{v,-}(\lb)\begin{pmatrix}1&0\\ie^{\frac{-2i}{\e}\phi^*(\lb;x)}&1\end{pmatrix}, &\mbox{as $\lb\in\Sigma_2\cap \mathcal V$}.
\end{align}
\item[(c)] In the double scaling limit where $\e\to 0$ and simultaneously $x\to x^-$ in such a way that $x-x^-=\bigO(\e^{2/3})$, $P_v$ satisfies the matching \begin{equation}P_v(\lb)=P^{(\infty)}(\lb)(I+o(1)), \qquad\mbox{ for $\lb\in\partial \mathcal V$.}\end{equation}
\end{itemize}

\subsubsection{Model RH problem related to Painlev\'e II}
We will construct the parametrix explicitly using a model RH problem for the so-called $\Psi$-functions associated to the Painlev\'e II equation, and in particular related to the Hastings-McLeod solution of Painlev\'e II.
This RH problem appeared several times in the literature, see e.g.\ \cite{FIKN, FN, BDJ, BI, CK}.
\subsubsection*{RH problem for $\Psi$}
\begin{itemize}
\item[(a)] $\Psi:\mathbb C \setminus (\Gamma_1 \cup \Gamma_2)\to \mathbb C^{2\times 2}$ is analytic, where
\[\Gamma_1=e^{i\frac{\pi}{6}}\mathbb R^+\cup e^{i\frac{5\pi}{6}}\mathbb R^+, \qquad \Gamma_2=e^{-i\frac{\pi}{6}}\mathbb R^+\cup e^{-i\frac{5\pi}{6}}\mathbb R^+,\] and $\Gamma_1, \Gamma_2$ are both oriented towards the right.
\item[(b)] $\Psi$ satisfies the following jump conditions on $\Gamma_1$ and $\Gamma_2$,
\begin{align}&\Psi_+(\zeta;s) = \Psi_-(\zeta;s)
    \begin{pmatrix} 1 & 0 \\ 1 & 1 \end{pmatrix},&\mbox{ as $\zeta\in\Gamma_1$},\\
&\Psi_+(\zeta;s) = \Psi_-(\zeta;s) \begin{pmatrix} 1 & -1 \\ 0 & 1
    \end{pmatrix},&\mbox{ as $\zeta\in\Gamma_2$.}
    \end{align}
\item[(c)] $\Psi(\zeta;s)
e^{i(\frac{4}{3}\zeta^3+s\zeta)\sigma_3}=I+\bigO(\zeta^{-1})$ as
$\zeta\to \infty$.
\end{itemize}

This RH problem is related to the Painlev\'e II equation in the following way.
Consider the linear system
\begin{align*}
 &
\frac{d}{d\zeta} \Psi(\zeta;s)
    = \begin{pmatrix}
     -4i\zeta^2-i(s+2q^2)  & 4\zeta q+2iy  \\
     4\zeta q -2i y   & 4i\zeta^2+i(s+2q^2)
    \end{pmatrix}
    \Psi(\zeta;s),\\
&
\frac{\partial}{\partial s}
    \Psi(\zeta;s)
     =  \begin{pmatrix}
     - i\zeta & q \\ q & i\zeta \end{pmatrix}
    \Psi(\zeta;s).
\end{align*}
The compatibility condition $\Psi_{\zeta s}=\Psi_{s\zeta}$ of this system implies that $q$ solves the Painlev\'e II equation (\ref{PII}) and that $y(s)=q'(s)$, and therefore this system is called the Lax pair associated to Painlev\'e II.
If we let $q$ be the Hastings-McLeod solution to Painlev\'e II, characterized by (\ref{HM1})-(\ref{HM2}), and $y=q'$, the RH problem for $\Psi$ can be solved using fundamental solutions to the Lax pair, see \cite{FN, FIKN}.
The RH problem is solvable for every complex value of $s$ at which $q(s)$ does not have a pole. It is known \cite{HastingsMcLeod} that $q$ is free of poles on the real line, so that the RH problem for $\Psi$ is solvable for $s$ in a neighborhood of $\mathbb R$. Condition (c) of the RH problem holds moreover uniformly for $s$ in compact subsets of $\mathbb C\setminus \mathcal P$, where $\mathcal P$ is the set of poles for $q$, and the asymptotic condition can be strengthened as follows,
\begin{equation}\label{RHP Psi: c2}
\Psi(\zeta;s)
e^{i(\frac{4}{3}\zeta^3+s\zeta)\sigma_3}=I+\frac{\Psi_1(s)}{\zeta}+\frac{\Psi_2(s)}{\zeta^2}+\bigO(\zeta^{-3}),\qquad\mbox{ as $\zeta\to\infty$,}
\end{equation}
with
\begin{equation}\label{Psi1}
\Psi_1(s)=\frac{1}{2i}\begin{pmatrix}p(s)&q(s)\\-q(s)&-p(s)
\end{pmatrix}, \qquad \Psi_2(s)=\frac{1}{2i}\begin{pmatrix}a(s)&ib(s)\\ib(s)&a(s)
\end{pmatrix},
\end{equation}
where
\begin{equation}\label{b}
p(s)=-q^4(s)-q^2(s)s+q'(s)^2, \qquad b(s)=\frac{q'(s)}{2}+\frac{1}{2}q(s)p(s),
\end{equation}
and $a(s)$ is unimportant for us.

\medskip

In order to create a RH problem with jumps which model the jumps which are necessary for the parametrix, we insert an additional parameter $\omega$ and
define
\begin{equation} \label{def-Phi}
\Phi(\zeta;s,\omega) =\sigma_1
   e^{i\omega \sigma_3}e^{-\frac{\pi i}{4}\sigma_3}\Psi(\zeta;s)e^{i(\frac{4}{3}\zeta^3+s\zeta)\sigma_3}e^{\frac{\pi i}{4}\sigma_3}
    e^{-i \omega \sigma_3}
    \sigma_1.
\end{equation}
We then have the following RH problem.

\subsubsection*{RH problem for $\Phi$}
\begin{itemize}
\item[(a)] $\Phi$ is analytic in
$\mathbb C \setminus (\Gamma_1 \cup \Gamma_2\cup\mathbb R)$.
\item[(b)] $\Phi$ satisfies the following jump conditions,
\begin{align}&\Phi_+(\zeta) = \Phi_-(\zeta)
    \begin{pmatrix} 1 &  ie^{-2i\omega}e^{2i(\frac{4}{3}\zeta^3+s\zeta)}\\ 0 & 1 \end{pmatrix},&\mbox{
     on $\Gamma_1$},\\
&\Phi_+(\zeta) = \Phi_-(\zeta) \begin{pmatrix} 1 & 0 \\ ie^{2i\omega}e^{-2i(\frac{4}{3}\zeta^3+s\zeta)} & 1
    \end{pmatrix},&\mbox{ on $\Gamma_2$.}
    \end{align}
\item[(c)] As $\zeta\to\infty$, we have
\begin{align}
&\label{RHP Phi: c}\Phi(\zeta;s,\omega)
=e^{-i\omega \sigma_3}\left(I+\frac{\Phi_1(s)}{\zeta}+\frac{\Phi_2(s)}{\zeta^2}
+O(\zeta^{-3})\right)e^{i\omega \sigma_3}
\end{align}
where
\begin{equation}
\Phi_1(s)=\frac{1}{2i}\begin{pmatrix}-p(s)&-iq(s)\\-iq(s)&p(s)
\end{pmatrix}, \qquad \Phi_2(s)=\frac{1}{2i}\begin{pmatrix}a(s)&-b(s)\\b(s)&a(s)
\end{pmatrix}
\end{equation}
This behavior is valid uniformly in $\omega$, and uniformly for $s$ in compact subsets of $\mathbb C\setminus\mathcal P$.
\end{itemize}

\subsubsection{Construction of the parametrix}

We construct $P_v$ explicitly in terms of the function $\Phi$.
Let us take $P_v$ of the following form,
\begin{equation}\label{def Pv}
    P_v(\lb;x,\e) = P^{(\infty)}(\lb) \Phi\left(\e^{-1/3} \zeta(\lb); \e^{-2/3} s(\lb;x), -\e^{-1}\phi(v;x)\right),
    \end{equation}
where $s(\lb;x)$ is analytic in $\overline{\mathcal V}$, and where $\zeta(\lb)$ is a conformal mapping in $\overline{\mathcal V}$ which maps $v$ to $0$. We can then specify the precise shape of the lenses $\Sigma_S$ near $v$ by requiring that $\zeta(\Sigma)\subset\Gamma_1\cup\Gamma_2$, so that $P$ has its jumps on $\Sigma_S$.

\medskip

We will now determine $\zeta(\lb)$ and $s(\lb;x)$
in such a way that
\begin{equation}\label{def phi f s}
\phi(\lb;x)-\phi(v;x)=\frac{4}{3}\zeta(\lb)^3+s(\lb;x)\zeta(\lb).
\end{equation}
If this condition is satisfied and if we let
\begin{equation}
\label{omega}
\omega(x,\e)=-\frac{1}{\e}\phi(v;x),
\end{equation}
the jump conditions (\ref{jump P1})-(\ref{jump P3}) for $P_v$ are valid using the jump conditions for $\Phi$ and analyticity of $P^{(\infty)}$ near $v$.
We can organize $\zeta$ and $s$ such that (\ref{def phi f s}) is valid, if we first define $\zeta$ by
\begin{equation}\label{def f}
\phi(\lb;x^-)-\phi(v;x^-)=\frac{4}{3}\zeta(\lb)^{3},
\end{equation}
so that using (\ref{phi1}), we have
\begin{equation}\label{f}
\zeta(v)=0, \qquad
\zeta'(v)=\frac{c^{1/3}}{2}, \qquad \zeta''(v)=\zeta'(v)\left(\dfrac{\frac{\partial^3}{\partial v^3}\theta(v;u)}{6\frac{\partial^2}{\partial v^2}\theta(v;u)}-\frac{1}{4(u-v)}\right),
\end{equation}
where $c$ has been defined in (\ref{phi2}) and $\theta(v;u)$ in (\ref{theta}).
Next define $s(\lb;x)$ by
\begin{equation}\label{def s}
s(\lb;x)\zeta(\lb)=\phi(\lb;x)-\phi(\lb;x^-)+\phi(v;x^-)-\phi(v;x)=(x-x^-)(\sqrt{u-\lb}-\sqrt{u-v}),
\end{equation}
so that we have
\begin{align}
&\label{s0}
s(v;x)=-\frac{x-x^-}{2\zeta'(v)\sqrt{u-v}}=-\frac{x-x^-}{c^{1/3}\sqrt{u-v}},\\
&\label{s1} s'(v;x)=s(v;x)\left(\dfrac{1}{4(u-v)}-\dfrac{1}{2}\dfrac{\zeta''(v)}{\zeta'(v)}\right).
\end{align}
Let us now take the double scaling limit where $\e\to 0$ and at the same time $x\to x^-$ in such a way that $\e^{-2/3}(x-x^-)\to X\ \in\mathbb R$. This means that
\begin{equation}\label{lims}
\e^{-2/3}s(v;x)\to s:=-\frac{1}{c^{1/3}\sqrt{u-v}}X.
\end{equation}
In this double scaling limit where in addition we let $\lambda$ approach $v$, we also have that
\[\lim \e^{-2/3}s(\lb;x)=s+\bigO(\lb-v).\] This implies that $\e^{-2/3}s(\lb;x)$ does not meet any poles of $q$ and lies in compact subsets of $\mathbb C\setminus \mathcal P$ for $\lb$ in a sufficiently small neighborhood $\mathcal V$ of $v$.
Summing up (\ref{def f}) and (\ref{def s}) now gives indeed (\ref{def phi f s}).

\medskip

If $\lb$ is at a fixed small distance away of $v$, for example on $\partial\mathcal V$, we have that $\e^{-1/3} \zeta(\lb)\to\infty$ if $\e\to 0$. Now we can use the asymptotic expansion (\ref{RHP Phi: c}) for $\Phi$.
Inserting this into (\ref{def Pv}), we obtain the following matching in the double scaling limit for $\lb\in \partial \mathcal V$,
\begin{multline}\label{matching P v}
P(\lb;x,\e)=P^{(\infty)}(\lb)e^{-i\omega\sigma_3}\\
\times\quad \left[I+\frac{\e^{1/3}}{\zeta(\lb)}\Phi_1(\e^{-2/3}s(\lb;x))+\frac{\e^{2/3}}{\zeta(\lb)^2}\Phi_2(\e^{-2/3}s(\lb;x))
+\bigO(\e)\right]e^{i\omega\sigma_3}.
\end{multline}

\subsection{Final transformation $S\mapsto R$}
Define
\begin{equation}
R(\lb;x,\e)=\begin{cases}
S(\lb)P_u(\lb)^{-1}, &\mbox{ in $\mathcal U$,}\\
S(\lb)P_v(\lb)^{-1}, &\mbox{ in $\mathcal V$,}\\
S(\lb)P^{(\infty)}(\lb)^{-1}, &\mbox{ elsewhere.}
\end{cases}
\end{equation}

Using the jump conditions for $S$ and the ones for the parametrices, one verifies that the jump matrices for $R$ are uniformly close to the identity matrix in the double scaling limit. This will imply that $R$ itself is also uniformly close to the identity matrix. In particular we have RH conditions for $R$ as follows.

\subsubsection*{RH problem for $R$}
\begin{itemize}
\item[(a)] $R$ is analytic in $\mathbb C\setminus \Sigma_R$, where $\Sigma_R=\Sigma_S\cup \partial \mathcal U \cup\partial \mathcal V$. Here we choose the clockwise orientation for $\partial \mathcal U$ and $\partial\mathcal V$.
\item[(b)] $R$ satisfies the jump condition $R_+(z)=R_-(z)v_R(z)$ on $\Sigma_R$, where $v_R$ has the following asymptotics in the double scaling limit,
\begin{equation}\label{asymptotics vr}
v_R(z)=\begin{cases}I+V_{1}(\lb;x,\e)\e^{1/3}+V_{2}(\lb;x,\e)\e^{2/3}+\bigO(\e), &\mbox{ as $\lb\in\partial \mathcal V\cup\partial \mathcal U$,}\\
I+\bigO(\e), &\mbox{ as $\lb\in (\mathcal U\cup \mathcal V)\cap \Sigma_R$,}\\
I+\bigO(e^{-c_1/\e}), &\mbox{ as $\lb\in \Sigma_R\setminus \overline{\mathcal U\cup \mathcal V}$.}
\end{cases}
\end{equation}
\item[(c)] $R(\lb )=I+\bigO(\lb^{-1})$ as $\lb\to\infty$.
\end{itemize}
The uniform decay of the jump matrix outside $\mathcal U\cup\mathcal V$ relies on Proposition \ref{prop jumps} and the jumps for the outside parametrix. The jumps inside $\mathcal U$ and $\mathcal V$ are easily verified to be $I+\bigO(\e)$ using (\ref{kappa3}), the jumps for $S$, and the jumps for $P$. The most important jumps (i.e.\ the only ones which will contribute to the expansion (\ref{expansionu}) for $u(x,t,\e)$) are the ones on $\partial\mathcal U$ and $\partial \mathcal V$.
Using (\ref{matching P u}) and (\ref{def Pinfty}) one verifies that, for $\lb\in\partial \mathcal U$, we have
\begin{equation}
\begin{split}
\label{v1u}V_{1}(\lb;x,\e)&=-\frac{is(\lb;x)^2}{4\e^{4/3}(-\zeta(\lb))^{1/2}}P^{(\infty)}(\lb)\sigma_3P^{(\infty)}(\lb)^{-1}\\
&=\frac{s(\lb;x)^2}{4\e^{4/3}(-\zeta(\lb))^{1/2}}
{\small \begin{pmatrix}
0&-(u-\lb)^{-1/2}\\
(u-\lb)^{1/2}&0
\end{pmatrix}},
\end{split}
\end{equation}
and for $\lb\in\partial\mathcal V$ we have using (\ref{matching P v}),
\begin{align}\label{v1v}&V_{1}(\lb;x,\e)=P^{(\infty)}(\lb)e^{-i\omega\sigma_3}\Phi_1(\e^{-2/3}s(\lb;x))e^{i\omega\sigma_3}P^{(\infty)}(\lb)^{-1}\nonumber \\
&=
\frac{p(\e^{-2/3}s(\lb;x))}{2 \zeta(\lambda)}
{\small \begin{pmatrix}0&(u-\lb)^{-1/2}\\-(u-\lb)^{1/2}&0
\end{pmatrix}}\nonumber \\
&\qquad\qquad
  -\frac{q(\e^{-2/3}s(\lb;x))}{2 \zeta(\lambda)} {\small \begin{pmatrix}\cos 2\omega(x,\e)&(u-\lb)^{-1/2}\sin 2\omega(x,\e)\\ (u-\lb)^{1/2}\sin 2\omega(x,\e) & -\cos 2\omega(x,\e)
\end{pmatrix}}
.
\end{align}
Note that $s(\lb;x)=\bigO(\e^{2/3})$, so that $V_1$ remains bounded in the double scaling limit.

\medskip

Let us now take a look at $V_2$. On $\partial\mathcal U$, we find the following by (\ref{matching P u}) and (\ref{def Pinfty}),
\begin{equation}
\label{v2u}V_{2}(\lb;x,\e)=-\frac{s(\lb;x)}{4\e^{2/3}\zeta(\lb)}P^{(\infty)}(\lb)\sigma_1
P^{(\infty)}(\lb)^{-1}
=-\frac{s(\lambda;x)}{4\e^{2/3}\zeta(\lb)}\sigma_3,
\end{equation}
and on $\partial\mathcal V$,
\begin{multline}
\label{v2v}V_{2}(\lb;x,\e)=P^{(\infty)}(\lb)M_2P^{(\infty)}(\lb)^{-1}\\
=\frac{a(\e^{-2/3}s(\lb;x))}{2i\zeta(\lb)^2}I\\ -\frac{b(\e^{-2/3}s(\lb;x))}{2\zeta(\lb)^2}{\small \begin{pmatrix}-\sin 2\omega(x,\e)&(u-\lb)^{-1/2}\cos 2\omega(x,\e)\\(u-\lb)^{1/2}\cos 2\omega(x,\e) & \sin 2\omega(x,\e)\end{pmatrix}}.
\end{multline}
Note that $V_1$ and $V_2$ can be extended to meromorphic functions in $\mathcal U\cup \mathcal V$, where $V_1$ has simple poles at $u,v$, and $V_2$ has a simple pole at $u$ and one of order two at $v$.
$V_1$ and $V_2$ remain bounded in the double scaling limit, but they do depend on $\e$.

\section{Asymptotics for $R$ and the KdV solution $u(x,t,\e)$}\label{section: final}

Because the jump matrix for $R$ has an asymptotic expansion in powers of $\e^{1/3}$, it follows \cite{Deift, DKMVZ1} that $R$ itself has a similar expansion (uniform in $\lambda$) in the double scaling limit,
\begin{equation}\label{Rexpansionepsilon}
R(\lambda;x,\e)=I+\epsilon^{1/3}R^{(1)}(\lb;x,\e)+\epsilon^{2/3}R^{(2)}(\lb;x,\e)+\bigO(\epsilon).
\end{equation}
We will compute large $\lambda$ asymptotics for $R^{(1)}$ and $R^{(2)}$ using the following proposition.

\begin{proposition}
As $\lambda\to\infty$, we have
\begin{align}
&\label{R1}R^{(1)}(\lb;x,\e)=\frac{1}{\lambda}\left(\Res(V_{1};v)+\Res(V_{1};u)\right)+\bigO(\lb^{-2}),\\
&\label{R2}R^{(2)}(\lb;x,\e)=\frac{1}{\lambda}\left(\Res(W;v)+\Res(W;u)\right)+\bigO(\lb^{-2}),
\end{align}
where
\begin{equation}\label{def W}
W(\lambda)=R^{(1)}(\lambda)V_{1}(\lambda)+V_{2}(\lambda).
\end{equation}
\end{proposition}
\begin{proof}
Combining (\ref{asymptotics vr}) with (\ref{Rexpansionepsilon}) gives us
\begin{multline*}\left(I+\epsilon^{1/3}R^{(1)}(\lb)+\epsilon^{2/3}R^{(2)}(\lb)\right)_+=
\left(I+\epsilon^{1/3}R^{(1)}(\lb)+\epsilon^{2/3}R^{(2)}(\lb)\right)_-\\
\times \left(I+V_{1}(\lb)\e^{1/3}+V_{2}(\lb)\e^{2/3}\right)+\bigO(\e), \qquad\mbox{ for $\lambda\in\partial \mathcal V\cup \partial \mathcal U$}.\end{multline*}
Collecting the terms of order $\bigO(\e^{1/3})$ and of order $\bigO(\e^{2/3})$ leads to the identities
\begin{align*}
&R_+^{(1)}(\lb)=R_-^{(1)}(\lb)+V_1(\lb), &\mbox{ for $\lambda\in\partial \mathcal U\cup \partial \mathcal V$,}\\
&R_+^{(2)}(\lb)=R_-^{(2)}(\lb)+R_-^{(1)}(\lb)V_1(\lb)+V_2(\lb), &\mbox{ for $\lambda\in\partial \mathcal U\cup \partial \mathcal V$.}
\end{align*}
Together with analyticity of $R^{(1)}$ and $R^{(2)}$, and the asymptotic condition
\[R^{(j)}(\lb)\to 0, \qquad\mbox{ as $\lb\to\infty$, $j=1,2$},\]
the above jump conditions constitute a uniquely solvable RH problem for $R^{(1)}$ and $R^{(2)}$.
One verifies that, since $V_1$ has simple poles at $u,v$, $R^{(1)}$ is given by
\begin{equation}\label{R1proof}
R^{(1)}(\lb)=
\begin{cases}\frac{1}{\lambda-v}\Res(V_1;v)+\frac{1}{\lambda-u}\Res(V_1;u)&\mbox{ for $\lb\in\mathbb C\setminus(\mathcal U\cup \mathcal V)$,}\\
\frac{1}{\lambda-v}\Res(V_1;v)+\frac{1}{\lambda-u}\Res(V_1;u)-V_1(\lambda)&\mbox{ for $\lb\in \mathcal U\cup \mathcal V$.}
\end{cases}
\end{equation}
Since $V_2$ has a simple pole at $u$ and a double pole at $v$, we have
\begin{equation}\label{R2proof}
R^{(2)}(\lb)=
\begin{cases}F(\lb)&\mbox{ for $\lb\in\mathbb C\setminus(\mathcal U\cup \mathcal V)$,}\\
F(\lb)-W(\lb)&\mbox{ for $\lb\in \mathcal U\cup \mathcal V$,}
\end{cases}
\end{equation}
with
\[F(\lb)=\frac{1}{\lambda-v}\Res(W;v)+\frac{1}{\lambda-u}\Res(W;u)+\frac{1}{(\lambda-v)^2}\Res((\lambda-v)W;v).\]
This implies (\ref{R1}) and (\ref{R2}).
\end{proof}
As $\lb\to\infty$, $R$ can be written as follows,
\begin{equation}\label{Rinfty}
R(\lb;x,\e)=I+\frac{R_1(x,\e)}{\lambda}+\bigO(\lambda^{-2}).
\end{equation}
Using the fact that $S(\lb)=R(\lb)P^{(\infty)}(\lb)$ for $\lb$ away from $u$ and $v$, using (\ref{uS}) and (\ref{RHP Pinfty c2}) we get the following identity,
\begin{equation}\label{uR}
u(x,t,\epsilon)=u-2\epsilon\partial_x
R_{1,12}(x,t,\epsilon).
\end{equation}
As a consequence of (\ref{R1})-(\ref{R2}), it follows that
\begin{equation}\label{R1expansion}
R_1=\left(\Res(V_{1};v)+\Res(V_{1};u)\right)\e^{1/3}+\left(\Res(W;v)+\Res(W;u)\right)\e^{2/3}+\bigO(\e)
\end{equation}
in the double scaling limit.
For notational convenience, let us write
$q:=q(\e^{-2/3}s(v;x))$, and similarly for $a, b, p, q'$ in the remaining part of the paper.
In what follows, we carefully analyze the residues in (\ref{R1expansion}) and the $x$-derivatives of their $12$-entries.
\medskip

\begin{proposition}\label{prop V1v}
In the double scaling limit, we have
\begin{align*}
{\rm (i) }& \ \e^{4/3}\partial_x \Res(V_{1,12};u)=-\frac{x-x^-}{2(6t+f'_L(u))},\\
{\rm (ii) }& \ \e^{4/3}\partial_x \Res(V_{1,12};v)=\frac{2q\cos( 2\omega)}{c^{1/3}}\e^{1/3}+\frac{q^2}{c^{2/3}(u-v)}\e^{2/3}+\frac{q'\sin( 2\omega)}{c^{2/3}(u-v)}\e^{2/3},\\
{\rm (iii) }& \ \e^{5/3}\partial_x \Res(W_{12};u)=\frac{q\sin(2\omega)(x-x^-)^2}{2\sqrt{u-v}c^{1/3}(6t+f_L'(u))}\e^{-2/3}+\bigO(\e),\\
{\rm (iv) }& \ \e^{5/3}\partial_x \Res(W_{12};v)=-\frac{q^2\cos(4\omega)\e^{2/3}}
{c^{2/3}(u-v)}  +\frac{2q\sin(2\omega)\e^{2/3}}{c^{2/3}}\left(\frac{2b}{q}\left[\frac{c_2}{6}-\frac{1}{4(u-v)}\right]\right. \\ & \left. -\frac{p+\frac{q'}{2q}}{u-v}+\left[\frac{c_2}{12}-\frac{3}{8(u-v)}\right]\frac{(x-x^-)^2\e^{-4/3}}{c^{2/3}(u-v)}
+\frac{c^{1/3}(x-x^-)^2\e^{-4/3}}{4\sqrt{u-v}(6t+f_L'(u))}\right)\\&\hspace{11,4cm}+\bigO(\e),
\end{align*}
with $\omega$ given by (\ref{omega}), $c$ by (\ref{def c}), and $c_2=\frac{\frac{\partial^3}{\partial v^3}\theta(v;u)}{\frac{\partial^2}{\partial v^2}\theta(v;u)}$.
\end{proposition}
\begin{proof}
\begin{itemize}
\item[(i)]
The first equation follows directly from (\ref{v1u}) using (\ref{zu}) and (\ref{su}).
\item[(ii)] Note first, using (\ref{omega}) and (\ref{s0}) that
\[\partial_x \omega(x;\e)=-\sqrt{u-v}\e^{-1}, \qquad \partial_x s(v;x)=-\frac{1}{c^{1/3}\sqrt{u-v}}.\]
Using these identities and taking the derivative of the residue of the $12$-entry in (\ref{v1v}), we find equation (ii) using (\ref{f}).
\item[(iii)] By (\ref{def W}), we have
\begin{equation}\label{W}
\Res(W;y)_{12}=R_{11}^{(1)}(y)\Res(V_1;y)_{12}+R_{12}^{(1)}(y)\Res(V_1;y)_{22}+\Res(V_2;y)_{12},
\end{equation}
both for $y=u$ and $y=v$.
For $y=u$, we have that $\Res(V_1;u)_{22}$ and $\Res(V_2;y)_{12}$ vanish. The only nonzero term is thus the first one. Since we know that $R_{11}^{(1)}(u)=\Res(V_1;v)_{11}/(u-v)$ by (\ref{R1proof}),
we conclude from (\ref{v1u}) and (\ref{v1v}) that
\begin{equation}
\partial_x\Res(W;u)_{12}=-\dfrac{q\sin(2\omega)s(u;x)^2}{4\sqrt{u-v}\zeta'(v)\sqrt{\zeta'(u)}}\e^{-2/3}+O(\e).
\end{equation}
Inserting the formulas (\ref{zu}), (\ref{su}), and (\ref{f}), we obtain (iii).
\item[(iv)] Taking a look at (\ref{W}) for $y=v$, all three terms are nonzero.
The third term at the right hand side of (\ref{W}) gives by (\ref{v2v}),
\begin{equation}\label{iv3}
\Res(V_2;v)_{12}=\dfrac{\cos(2\omega)}{2\zeta'(v)^2\sqrt{u-v}}
\left( \dfrac{b\zeta''(v)}{\zeta'(v)}-b's'(v;x)\e^{-2/3}-\dfrac{b}{2(u-v)}\right).
\end{equation}
For the first two terms in (\ref{W}), we use (\ref{R1proof}) and obtain
\[
R^{(1)}_{11}(v)=\dfrac{\cos(2\omega)}{2\zeta'(v)}
\left(q's'(v,x)\epsilon^{-2/3}-\dfrac{q}{2}\dfrac{\zeta''(v)} {\zeta'(v)}\right),
\]
\begin{multline*}
R^{(1)}_{12}(v)=\dfrac{1}{2\zeta'(v)\sqrt{u-v}}\left(\dfrac{p\zeta''(v)}{2\zeta'(v)}-p's'(v;x)\e^{-2/3}-
\dfrac{p}{2(u-v)}\right)\\
\qquad\qquad\qquad -\dfrac{\sin(2\omega)}{2\zeta'(v)\sqrt{u-v}}\left(\dfrac{q\zeta''(v)}{2\zeta'(v)}-q's'(v;x)\e^{-2/3}-
\dfrac{q}{2(u-v)}\right)\\
-\dfrac{s^2(u;x)\e^{-4/3}}{4(u-v)\sqrt{\zeta'(u)}}.
\end{multline*}
Now we need to use the fact that $2b=q'+qp$ and that $p'=-q^2$ (see (\ref{b}) and (\ref{PII})).
After a straightforward but somewhat lengthy calculation, using the above equations together with (\ref{v1u})-(\ref{v1v}) and (\ref{W}), we find
\begin{multline*}
\Res(W_{12},v)=\dfrac{q^2\sin(4\omega)}{16\zeta'(v)^2(u-v)^{3/2}}+\dfrac{q\cos(2\omega)}{4\zeta'(v)^2\sqrt{u-v}}\\
\times \ \left(\dfrac{2b\zeta''(v)}{q\zeta'(v)}-\dfrac{p+\frac{q'}{2q}}{u-v}-s'(v;x)s(v;x)\e^{-4/3}-
\dfrac{\zeta'(v)s(u;x)^2}{2\sqrt{u-v}\sqrt{\zeta'(u)}}\e^{-4/3}\right).
\end{multline*}
Taking derivatives and using (\ref{zu}), (\ref{su}), (\ref{f}), and (\ref{s0})-(\ref{s1}), (iv) is obtained directly.
\end{itemize}
\end{proof}

Now we can insert the above estimates into (\ref{R1expansion}) and (\ref{uR}). However we need to be careful with the derivative of the $\bigO(\e)$-term in (\ref{R1expansion}). On $\partial\mathcal U$, one can check that this term is of order $\bigO(\e)$ by computing the next term in the expansion (\ref{matching P u}). On $\partial\mathcal V$, using (\ref{matching P v}), we have a contribution of order $\bigO(\omega\e)=\bigO(1)$. Also on the other parts of the contour, we have a contribution of order $\bigO(\e)$.
Taking this into account, we obtain
{\small \begin{multline}\label{uR2}
u(x,t,\epsilon)=u(t)-4\dfrac{q\cos (2\omega)}{c^{1/3}}\e^{1/3}-\dfrac{4q^2\sin^2(2\omega)}{c^{2/3}(u-v)}\e^{2/3}+\frac{x-x^-}{6t+f'_L(u)}\\
-\frac{4q\sin(2\omega)\e^{2/3}}{c^{2/3}}\left(\frac{2b}{q}\left[\frac{c_2}{6}-\frac{1}{4(u-v)}\right] -\frac{p+\frac{q'}{2q}}{u-v}+\left[\frac{c_2}{12}-\frac{3}{8(u-v)}\right]\frac{(x-x^-)^2\e^{-4/3}}{c^{2/3}(u-v)}\right.
\\
\left. +\frac{c^{1/3}(x-x^-)^2\e^{-4/3}}{2\sqrt{u-v}(6t+f_L'(u))}\right)+\bigO(\e).
\end{multline}
}
Observe that $2\omega=-\frac{1}{\e}\Theta(x,t)$, with $\omega$ defined by (\ref{omega}) and $\Theta$ by (\ref{Theta}).
After a brief calculation, we can now write (\ref{uR2}) in the form (\ref{expansionu}), so that Theorem~\ref{theorem: main} is proven.

\section*{Acknowledgements}
The authors are grateful to Ken McLaughlin for useful remarks.
TC is a Postdoctoral Fellow of the Fund for Scientific Research - Flanders (Belgium), and was also supported by
Belgian Interuniversity Attraction Pole P06/02, FWO-Flanders project G042709, K.U.Leuven research grant OT/08/33, and by ESF program MISGAM.
TG  acknowledges support by the ESF program MISGAM, by the RTN ENIGMA and by Italian
COFIN 2004 ``Geometric methods in the theory of nonlinear waves and their applications''.

\obeylines
\texttt{
Tom Claeys
Department of Mathematics
Katholieke Universiteit Leuven
Celestijnenlaan 200B
B-3001 Leuven, BELGIUM
E-mail: tom.claeys@wis.kuleuven.be
\bigskip

Tamara Grava
SISSA
Via Beirut 2-4
34014 Trieste, ITALY
E-mail: grava@sissa.it
}

\end{document}